\documentclass[sigconf]{acmart}
\usepackage{wrapfig}
\usepackage{float}
\usepackage{subcaption}
\usepackage[normalem]{ulem}

\usepackage{tikz}
\usetikzlibrary{shapes.geometric, arrows.meta, positioning}

\tikzstyle{block} = [rectangle, rounded corners, minimum width=3.2cm, minimum height=0.5cm,text centered, draw=black, fill=blue!10]
\tikzstyle{arrow} = [thick,->,>=stealth]

\usepackage{xcolor}

\copyrightyear{2025}
\acmYear{2025}
\setcopyright{cc}
\setcctype{by}
\acmConference[GECCO '25]{Genetic and Evolutionary Computation Conference}{July 14--18, 2025}{Malaga, Spain}
\acmBooktitle{Genetic and Evolutionary Computation Conference (GECCO '25), July 14--18, 2025, Malaga, Spain}\acmDOI{10.1145/3712256.3726415}
\acmISBN{979-8-4007-1465-8/2025/07}

\AtBeginDocument{%
  }

\begin{document}

\title{Evolving Hard Maximum Cut Instances for Quantum Approximate Optimization Algorithms}

\author{Shuaiqun Pan}
\orcid{0000000170394875}
\affiliation{%
  \institution{LIACS, Leiden University}
  \city{Leiden}
  \country{The Netherlands}}
\email{s.pan@liacs.leidenuniv.nl}

\author{Yash J. Patel}
\orcid{0009-0007-3060-6950}
\affiliation{%
  \institution{applied Quantum algorithms \&}
  \institution{LIACS, Leiden University}
  \city{Leiden}
  \country{The Netherlands}
}
\email{y.j.patel@liacs.leidenuniv.nl}

\author{Aneta Neumann}
\orcid{0000-0002-0036-4782}
\affiliation{%
  \institution{Optimisation and Logistics, School of Computer and Mathematical Sciences, The University of Adelaide}
  \city{Adelaide}
  \country{Australia}}
\email{aneta.neumann@adelaide.edu.au}

\author{Frank Neumann}
\orcid{0000-0002-2721-3618}
\affiliation{%
  \institution{Optimisation and Logistics, School of Computer and Mathematical Sciences, The University of Adelaide}
  \city{Adelaide}
  \country{Australia}}
\email{frank.neumann@adelaide.edu.au}

\author{Thomas B{\"a}ck}
\orcid{0000000167681478}
\affiliation{%
\institution{LIACS, Leiden University}
  \city{Leiden}
  \country{The Netherlands}}
\email{t.h.w.baeck@liacs.leidenuniv.nl}

\author{Hao Wang}
\orcid{0000000249335181}
\affiliation{%
  \institution{applied Quantum algorithms \&}
  \institution{LIACS, Leiden University}
  \city{Leiden}
  \country{The Netherlands}}
\email{h.wang@liacs.leidenuniv.nl}

\begin{abstract}
Variational quantum algorithms, such as the Recursive Quantum Approximate Optimization Algorithm (RQAOA), have become increasingly popular, offering promising avenues for employing Noisy Intermediate-Scale Quantum devices to address challenging combinatorial optimization tasks like the maximum cut problem. In this study, we utilize an evolutionary algorithm equipped with a unique fitness function. This approach targets hard maximum cut instances within the latent space of a Graph Autoencoder, identifying those that pose significant challenges or are particularly tractable for RQAOA, in contrast to the classic Goemans and Williamson algorithm. Our findings not only delineate the distinct capabilities and limitations of each algorithm but also expand our understanding of RQAOA's operational limits. Furthermore, the diverse set of graphs we have generated serves as a crucial benchmarking asset, emphasizing the need for more advanced algorithms to tackle combinatorial optimization challenges. Additionally, our results pave the way for new avenues in graph generation research, offering exciting opportunities for future explorations.
\end{abstract}



\maketitle

\section{Introduction}

In the era of noisy intermediate-scale quantum (NISQ) technology, the practical use of quantum computing is being increasingly pursued, despite challenges like the scarcity of quantum resources and the noise in quantum gates~\cite{preskill2018quantum}. Key research areas such as combinatorial optimization~\cite{farhi2014quantum}, quantum machine learning~\cite{benedetti2019parameterized}, and quantum chemistry~\cite{moll2018quantum} are recognized as potential fields where NISQ devices might surpass best classical methods, thereby showcasing a quantum advantage.

The Recursive Quantum Approximate Optimization Algorithm (RQAOA)~\cite{bravyi2020obstacles} was developed to improve upon the Quantum Approximate Optimization Algorithm (QAOA) for combinatorial optimization problems like maximum cut. While QAOA~\cite{farhi2014quantum} was initially introduced as a quantum heuristic, research indicates it struggles to surpass the Goemans and Williamson (GW) algorithm~\cite{goemans1995improved}, a classical method with strong approximation guarantees, at constant depth~\cite{hastings2019classical, farhi2020quantum, marwaha2021local, barak2021classical}. In contrast, RQAOA employs a recursive strategy that iteratively refines solutions, demonstrating superior performance both numerically and analytically~\cite{bravyi2021classical, bravyi2022hybrid, bae2024recursive}. This makes RQAOA a promising advancement in quantum optimization.


Research has shown that certain algorithms underperform in specific scenarios, often labeled as hard or worst-case instances~\cite{ko1986hard}. Yet, guidance on generating or identifying such challenging instances remains limited in the literature. For instance, the RQAOA paper highlights an infinite family of $d$-regular bipartite graphs where the QAOA lags behind the GW algorithm when the depth (number of layers) is $O(log \textit{n})$, where $\textit{n}$ is the number of nodes~\cite{bravyi2020obstacles}. However, the paper does not provide concrete methods for constructing such graphs. In our work, we aim to bridge this gap by leveraging evolutionary computation method to generate instances that reveal the contrasting behaviors of quantum and classical optimization algorithms, specifically RQAOA and GW, when solving the maximum cut problem. While various heuristics have been explored in~\cite{dunning2018works}, we select the GW algorithm due to its well-established approximation guarantees and extensive theoretical characterization, making it a suitable baseline for comparative analysis. From the evolutionary computation perspective, our contribution lies not in proposing a new variant of Covariance Matrix Adaptation Evolution Strategy (CMA-ES)~\cite{hansen2001completely, hansen2003reducing}, but in repurposing it for a novel problem setting: rather than optimizing solutions to fixed problems, we use CMA-ES to evolve the problem instances themselves. Specifically, we operate within the latent space of a well-trained Graph Autoencoder (GAE), using CMA-ES to search for latent representations that yield graphs which are intentionally easy for one algorithm but hard for the other. The fitness function is carefully designed to reward maximum divergence in performance between RQAOA and GW.


We deliver several significant contributions:
\begin{itemize}
    \item We introduce a method that effectively pinpoints maximum cut instances that either pose significant challenges or are considerably more manageable for the RQAOA in comparison to the GW algorithm.
    \item Through an extensive analysis of these instances across various sizes of the maximum cut problem, our generated instances serve as valuable benchmarks for evaluating and contrasting the effectiveness of different algorithms in addressing maximum cut challenges. This benchmarking not only underscores performance discrepancies among algorithms but also aids in advancing quantum computing methodologies.
    \item Our research brings a fresh perspective to graph generation studies within the network science community, spurring further investigation and innovation in this domain.
\end{itemize}

\section{PRELIMINARIES}
\label{sec:preliminaries}

\paragraph{Maximum cut}
The maximum cut problem in graphs is one of the classical and well-studied NP-hard combinatorial optimization problems. Given an undirected graph $G=(V,E)$, the goal is to compute a partitioning into two sets $S \subseteq V$ and $V \setminus S$ such that the number of edges 
$$C(S) = |\{e \in E \mid e \cap S \not = \emptyset \wedge e \cap (V \setminus S)  \not = \emptyset\}|$$ crossing the two sets is maximal.

\paragraph{Goemans and Williamson algorithm}
The maximum cut problem is APX-hard~\cite{DBLP:journals/jcss/PapadimitriouY91} and can be approximated using the GW algorithm~\cite{goemans1995improved}.
Specifically, this approach involves solving a relaxed version of the original problem. We first tackle an optimization problem in the continuous domain $[-1, 1]^N$ using a semidefinite program, which provides us with a real-valued solution associated with a cost denoted as $C_{\text{relaxed}}$. Then, a sampling process, guided by probabilities derived from the entries of the real-valued solution (via a random projection method), is used to produce a feasible solution in the form of a binary string.
It is important to note that the cost of the generated binary string is guaranteed to be at most equal to $C_{\text{relaxed}}$, and it is possible that $C_{\text{max}} < C_{\text{relaxed}}$. However, the expected value obtained by the cost function using this method is proven to be less than $\alpha C_{\text{max}}$, where $\alpha \approx 0.878$, for any input graph. In other words, this is an $\alpha$-approximation algorithm.

\begin{figure}
    \centering
     \includegraphics[width=\linewidth]{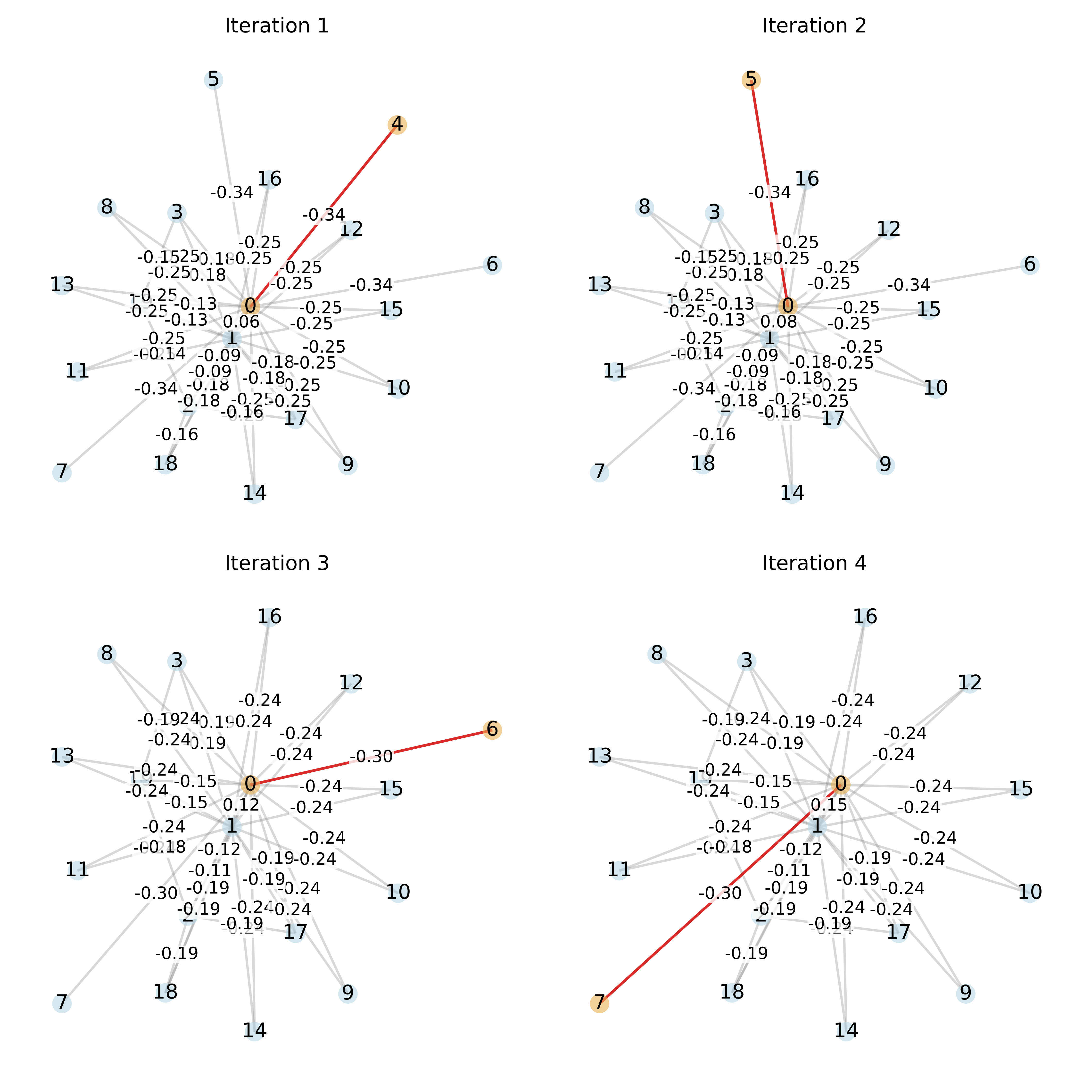}
     \caption{RQAOA decision process ($n_c = 10$) across the first four iterations on a 20-node maximum cut instance, where it surpasses the GW algorithm. Each subpanel represents one algorithm iteration, with nodes as blue dots and edges as grey lines. Edge labels show the expectation values derived from the quantum Hamiltonian (the quantum analog of the maximum cut cost function). At each step, the edge with the largest absolute expectation value is selected and contracted, highlighted in orange along with its endpoints, thus reducing the graph size by at least one node.}
     \label{fig:rqaoa_steps}
     \Description{Visualization of RQAOA performance ($n_c$ is set to 10) over the first four iterations for a 20-node maximum cut instance, where it outperforms the GW algorithm.}
\end{figure}

\paragraph{Recursive Quantum Approximate Optimization Algorithm}
The maximum cut problem has received significant attention in the area of quantum optimization and different quantum algorithms have been designed. 
As discussed earlier, QAOA faces significant limitations due to its intrinsic locality and symmetry constraints, with substantial theoretical research having established no-go theorems in this area. To address these challenges, RQAOA~\cite{bravyi2020obstacles} was introduced as an alternative quantum optimization algorithm specifically for combinatorial optimization problems. During each iteration of RQAOA, the problem size—represented as a graph or hypergraph—is reduced by one or more elements, based on correlations between variables derived from QAOA. This reduction process introduces non-local effects by creating new connections between nodes that were not previously linked, thus circumventing the locality constraints of QAOA.
Finally, the iterative process stops once the number of variables is below a certain predefined threshold $n_c$ ($n_c \approx O(1) \ll |V|$) such that a classical solver can be used to solve the remaining graph. 
In this study, we focus on depth-1 RQAOA, which can be classically simulated with a runtime of $O(n^4)$~\cite{patel2024reinforcement}.
Understanding the performance of these algorithms is challenging and pointing out instances where quantum approaches perform significantly better or worse than the classical GW algorithm provides an important understanding of their performance in comparison to state-of-the-art classical approximation algorithms.

Fig.~\ref{fig:rqaoa_steps} illustrates the initial decision-making process that RQAOA employs over four iterations, with $n_c=10$, on a generated 20-node maximum cut instance where RQAOA outperforms the GW algorithm. 
Each subfigure corresponds to one iteration, highlighting the single edge chosen for contraction.
The selection is based on the largest absolute expectation value under the quantum state derived from QAOA: a positive sign indicates that the edge should remain uncut, while a negative sign indicates that it should be in the cut.
In the event of a tie among these largest expectation values, any of the tied edges may be chosen arbitrarily.
The selected edge and its corresponding nodes are highlighted in orange. The visualization offers an intuitive look at the way RQAOA navigates the solution space using quantum-informed decisions, distinguishing its behavior from classical algorithms. 
However, note that in this work we consider RQAOA at depth-1, which can be simulated classically, making it less clear whether its decisions might be parallel to those of some classical heuristic.
However, at higher depths beyond classical simulability, RQAOA may have access to information not available to classical algorithms, offering a potential route to quantum advantage.


\paragraph{Features of Graphs} To effectively analyze the generated maximum cut instances, we compute a comprehensive suite of graph-level features that capture both intrinsic structural characteristics and algorithm-dependent behaviors. The structural features include classical topological indicators such as graph density, girth, chromatic number, assortativity, and transitivity, alongside spectral properties derived from the normalized Laplacian. Specifically, the spectral gap measures the difference between the top two eigenvalues and reflects how well-connected the graph is overall. The log-transformed largest and second-largest eigenvalues help identify structures like clusters or bottlenecks. Of particular interest is the log of the smallest non-zero eigenvalue, which indicates weak connectivity or the presence of nearly isolated nodes. We further include combinatorial descriptors like the normalized size of the maximum independent set to reflect sparsity and node connectivity. To evaluate how each instance interacts with the GW algorithm, we extract several features from its semidefinite programming (SDP) formulation, including the expected cut value, its variability across randomized rounding procedures, and statistics from the lower triangular portion of the SDP factor matrix. A complete list and detailed definitions of all features are presented in sections~\ref{AppendixA} and~\ref{AppendixB} (see the supplementary material~\cite{supplementary}).

\paragraph{Graph Neural Networks (GNNs)} GNNs have been proposed to tackle learning problems involving non-Euclidean data~\cite{DBLP:journals/corr/abs-2403-04468,zhu2022survey}, among which graph autoencoders (GAEs) are successfully applied to unsupervised tasks to learn graph topologies~\cite{KipfW16a}. GAEs encode vertices or the entire graph into a latent space, where several different encoding/decoding methods are devised. GAEs have been used to extract network representations and learn the empirical distributions thereof~\cite{wu2020comprehensive}. Due to the famous graph isomorphism issues, the generating/decoding component of GNNs is non-trivial: several works target this issue, including GraphRNN~\cite{you2018graphrnn} employing a sequential method to construct nodes and edges by learning from a representative set of graphs; GraphVAE~\cite{simonovsky2018graphvae} that uses variational autoencoders~\cite{kingma2013auto} to generate graphs from continuous embeddings.

\paragraph{Covariance Matrix Adaptation Evolution Strategy (CMA-ES)} CMA-ES~\cite{Hansen06, Hansen16a} is the state-of-the-art numerical algorithm for black-box optimization problems. CMA-ES employs a set/population of search points, which are sampled from a multivariate Gaussian distribution. CMA-ES selects the subset of points with better objective values and uses their information to update the mean and covariance of the Gaussian. CMA-ES has been successfully applied to various challenging applications and is known for its fast convergence and robustness~\cite{HansenNGK09, LoshchilovH16,bonet2023performance, SalimansHCS17}.

\section{Related work}
In optimization, generating diverse problem instances is essential for effective benchmarking. The choice of instances significantly impacts algorithm evaluation, making it crucial to update benchmarks regularly to prevent overfitting specific cases. Assessing benchmark quality is challenging, as measuring diversity and representativeness is not straightforward. Typically, evaluation considers key dimensions such as feature space, performance space, and instance space~\cite{bartz2020benchmarking}.

The literature highlights various methods for generating maximum cut instances used in benchmarking. A notable example is the Balasundarm-Butenko problem generator~\cite{balasundaram2005constructing}, which utilizes a framework designed for creating test functions within global optimization. This involves applying continuous formulations to combinatorial optimization problems. Another key tool is the machine-independent graph generator known as \emph{Rudy}, introduced by~\cite{helmberg2000spectral} which has been widely used in subsequent research to evaluate algorithms~\cite{benson1999mixed, burer2001projected, burer2002rank}. In 2018, a comprehensive library~\cite{dunning2018works} containing 3,296 diverse instances was introduced, aggregating data from multiple sources, and further enriching the resources available for algorithm testing.

Evolutionary algorithms (EAs) have proven effective in generating diverse problem instances across various domains. \citet{bossek2019evolving} pioneered an EA-based approach for crafting the Travelling Salesperson Problem (TSP) instances using novel mutation operations. \citet{DBLP:conf/gecco/NeumannG0019} developed a method for evolving diverse TSP instances and images using multi-objective indicators. \citet{gao2021feature} introduced an EA-driven framework to generate problem instances classified as easy or hard based on distinct features. Addressing diversity in high-dimensional feature spaces, \citet{marrero2023generating} proposed a Knapsack problem instance generator combining Principal Component Analysis with novelty search. More recently, research has explored evolving instances for the chance-constrained maximum coverage problem~\cite{sadeghi2024evolving}. This study departs from traditional approaches that rely on approximation ratios in the $(1+1)$ EA, instead incorporating performance ratio variances for improved evaluation.

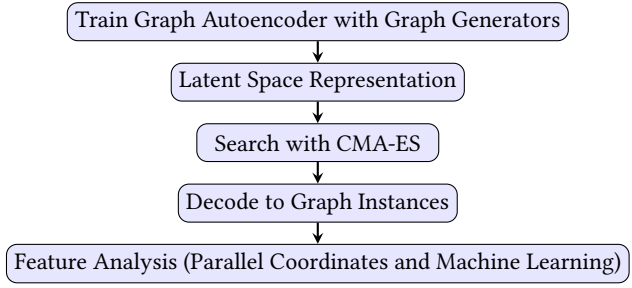
\begin{figure}[!ht]
\centering
\begin{tikzpicture}[node distance=0.8cm]

\node (gae) [block] {Train Graph Autoencoder with Graph Generators};
\node (latent) [block, below of=gae] {Latent Space Representation};
\node (cmaes) [block, below of=latent] {Search with CMA-ES};
\node (decode) [block, below of=cmaes] {Decode to Graph Instances};
\node (analysis) [block, below of=decode] {Feature Analysis (Parallel Coordinates and Machine Learning)};

\draw [arrow] (gae) -> (latent);
\draw [arrow] (latent) -> (cmaes);
\draw [arrow] (cmaes) -> (decode);
\draw [arrow] (decode) -> (analysis);

\end{tikzpicture}
\Description{Overview of the instance generation and evaluation pipeline.}
\caption{Overview of the instance generation and evaluation pipeline.}
\label{fig:pipeline}
\end{figure}

\section{Evolving Hard Maximum Cut Instances}


GNNs have proven effective for generating graphs with diverse attributes. In our research, we use tools like the GAE to explore and identify specific graph structures. Our goal is to identify hard maximum cut instances. We begin by training a GAE with graph generators that are rooted in network science. After the GAE is adequately trained, we proceed to pinpoint particular graph instances within its complex latent space. The identification process is facilitated by the classical CMA-ES algorithm. This strategy employs a fitness function to assess scenarios where the RQAOA either surpasses or falls short compared to the GW algorithm. Furthermore, we perform a comprehensive analysis of the generated instances using a parallel coordinate plot and a dedicated machine learning pipeline. The overall method of the instance generation and evaluation pipeline is illustrated in the flowchart shown in Fig.~\ref{fig:pipeline}.

\paragraph{Problem formulation}
We aim to propose a method for evolving hard instances of the maximum cut problem, to challenge a recently proposed quantum search algorithm - RQAOA, against a well-studied classical algorithm - GW. We measure the largest number of cuts found by each algorithm on a graph instance $I$, denoted as $\textit{p}_{\text{RQAOA}}(\text{I})$, or $\textit{p}_{\text{GW}}(\text{I})$.

\paragraph{Graph instance representation}
We choose to base our research on Permutation-Invariant Variational Autoencoder (PIGVAE)~\cite{DBLP:conf/nips/WinterNC21} as it handles the graph isomorphism problem explicitly.
The core principle of PIGVAE is that identical graphs should yield the same representation, regardless of the order of their nodes. It combines a variational autoencoder with a permuter model. This permuter model, trained in conjunction with the encoder and decoder, assigns a permutation matrix to each input graph. This matrix aligns the node order of the input graph with that of its reconstruction, ensuring consistent representation across different inputs.

\paragraph{Evolving instances with CMA-ES}
We aim to search hard graph instances for a maximum cut solver against another solver. Hence, we propose using the following objective function: 
\begin{equation}
\textit{p}_{\text{B}}(\text{I}) / \textit{p}_{\text{A}}(\text{I}),
\end{equation}
where $A$ is the baseline/reference algorithm and $B$ is the one of interest. In this paper, we try to answer two questions: is there a cluster/family of graphs (1) hard to solve by RQAOA w.r.t.~the classical GW method; (2) vice versa? For those two questions, we maximize the objective functions, $\textit{p}_{\text{GW}}(\text{I}) / \textit{p}_{\text{RQAOA}}$ and $\textit{p}_{\text{RQAOA}}(\text{I}) / \textit{p}_{\text{GW}}$, respectively, in the latent space of the already-trained PIGVAE model. We utilize the widely applied CMA-ES for this maximization task. For each latent point sampled by the CMA-ES, we first decode the latent point to a graph instance $I$ and then execute RQAOA and GW algorithms on instance $I$ 100 times due to their stochasticity. We take the median value of each algorithm to compute the above ratio.
Importantly, a randomly sampled latent point may be decoded into a disconnected graph. We might initiate the CMA-ES with connected graphs. Still, as the search progresses, CMA-ES may sample disconnected graphs again. To resolve this issue, we propose the following strategy: for any latent point that corresponds to a disconnected graph, we solve the maximum cut problems supported on each connected component and then take the sum of the number of cuts from all components.
For RQAOA, if the number of nodes of a component is smaller than a preset value $\textit{n}_c$, we solve the maximum cut problem on this component by a brute force method; Otherwise, we apply RQAOA on it. In the recursion step of RQAOA, disconnectedness can be created even if we start with a connected graph. In this case, we apply the above strategy to the connected components.
In the experiments, we test our approach with two sizes: 20 nodes and 100 nodes, and we set $\textit{n}_c$ to 10 and 20, respectively.

\paragraph{Feature analysis} To better understand the structural factors that drive differences in algorithmic performance, we perform an in-depth feature analysis of the generated maximum cut instances. This analysis combines two complementary strategies: visualization through parallel coordinate plots and predictive modeling via a machine learning pipeline. We examine instances generated from multiple CMA-ES runs, focusing on both 20-node and 100-node graphs where RQAOA or the GW algorithm demonstrates superior performance. Each instance is described using 18 features, which include both general graph-theoretic properties and metrics derived from the GW algorithm's semidefinite relaxation. The visualizations reveal characteristic trends in feature distributions tied to algorithmic success, while an AutoML-based classification framework, \texttt{TPOT} library~\cite{DBLP:conf/gecco/OlsonBUM16}, is used to train models capable of predicting the dominant algorithm for a given instance.

\section{Experiments}

\subsection{Experimental setup}

\begin{figure}[!ht]
    \centering
    \includegraphics[width=\linewidth]{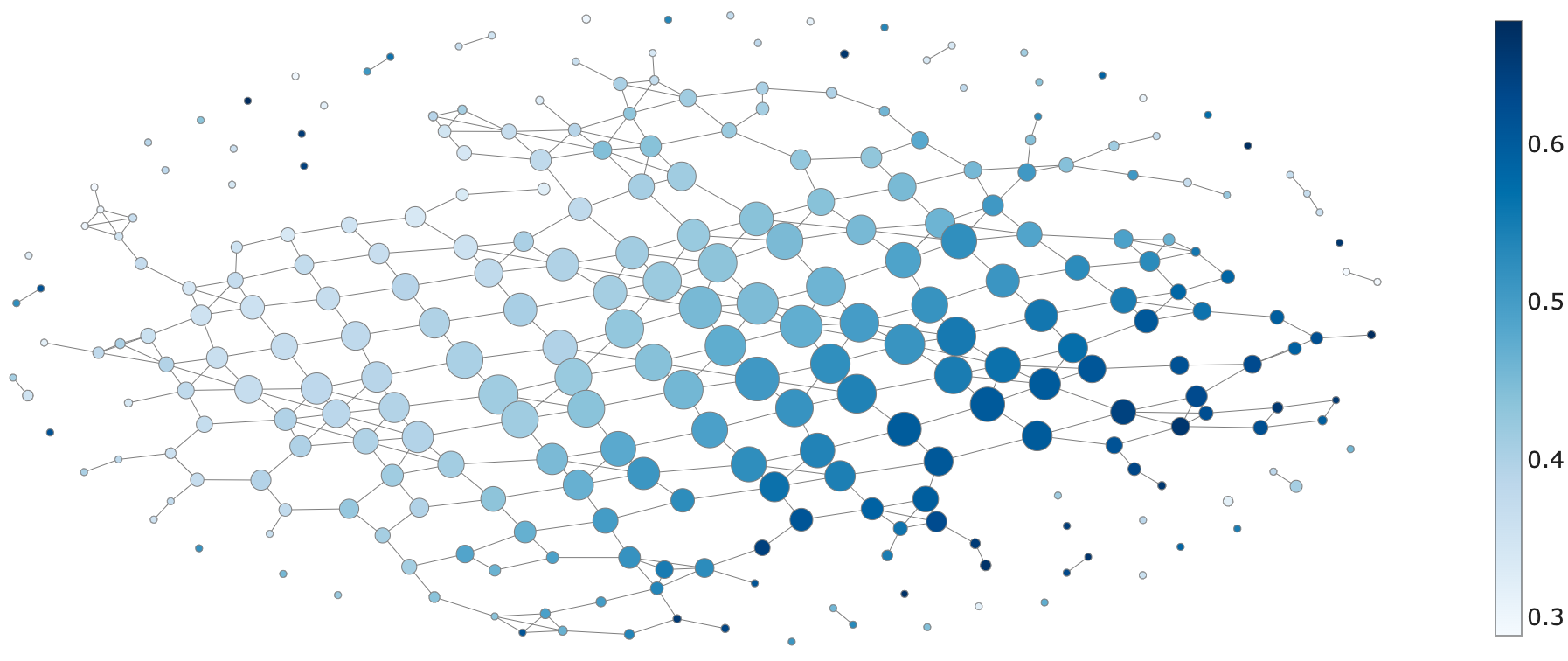}
    \includegraphics[width=\linewidth]{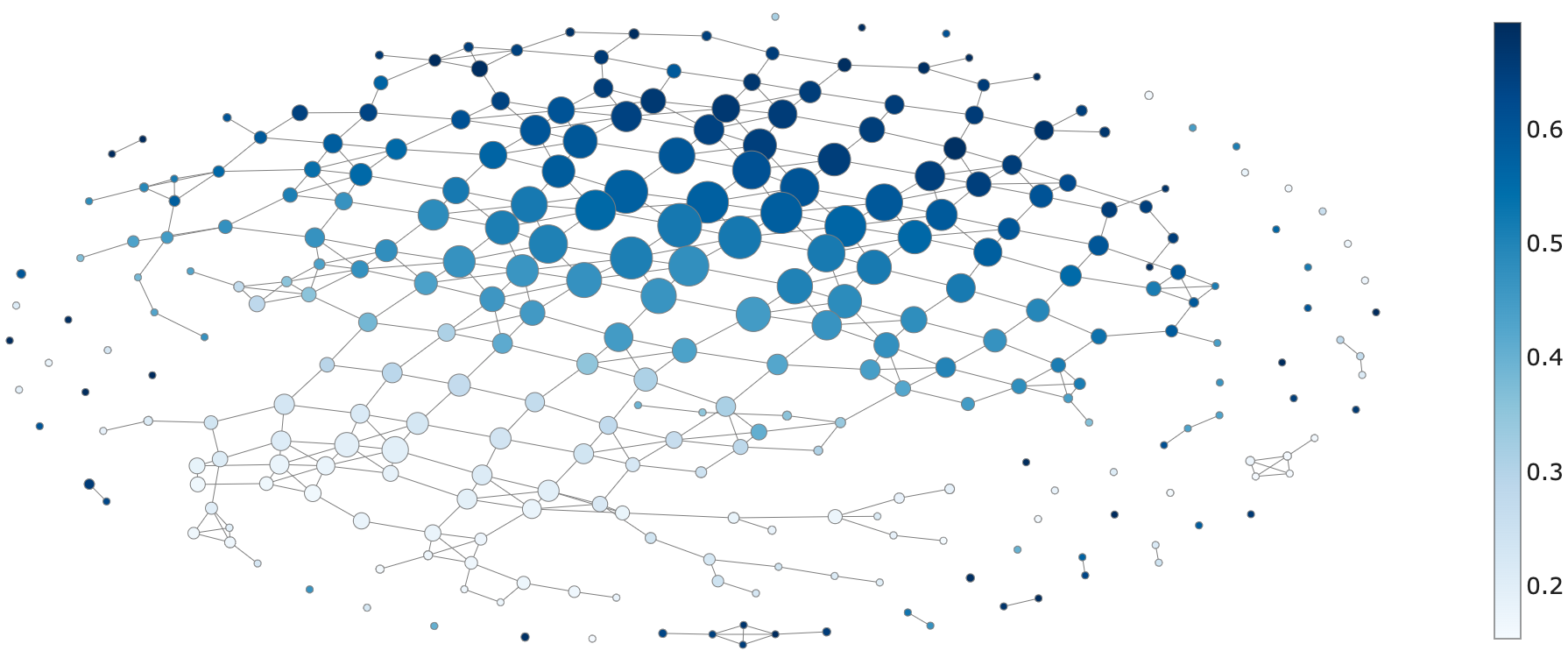}
    \caption{Log-transformed second-largest eigenvalues of the normalized graph Laplacian for graph embeddings of 20-node (top) and 100-node (bottom) graphs. Nodes represent DBSCAN-identified clusters~\cite{ester1996density, schubert2017dbscan}, with edges indicating feature overlap or transitions between clusters. Node colors reflect the second-largest eigenvalue, offering insights into graph connectivity and structural complexity.}
    \Description{Second-largest eigenvalues of the normalized graph Laplacian for graph embeddings of 20-node (top) and 100-node (bottom) graphs.}
    \label{fig:20_100node_log__secondlargesteigval}
\end{figure}

\paragraph{GAE training}
We sample the training graph instance u.a.r.~from six random graph models: Erdős–Rényi~\cite{erdos1960evolution}, Watts-Strogatz small-world~\cite{watts1998collective}, Newman-Watts-Strogatz small-world~\cite{newman1999renormalization}, Random Regular~\cite{steger1999generating}, Barabási–Albert~\cite{barabasi1999emergence}, and Dual-Barabási–Albert graphs~\cite{moshiri2018dual}. We also take the parameter setting of those graph models and hyperparameters of the PIGVAE model from~\cite{DBLP:conf/nips/WinterNC21}. In each training epoch, we generate 6\,000 graph instances from the above random models.
We train the PIGVAE model with maximally 3\,000 epochs and a batch size of 128. An early stopping mechanism is implemented to terminate training if the validation loss shows no improvement over the past 60 epochs. 
To assess our approach on different graph sizes, we train two separate models: one for 20-node instances, trained using a single A100 GPU, and the other for 100-node instances, trained using four A100 GPUs.


\paragraph{CMA-ES setup}
CMA-ES operates in the latent space of the GAE. The search space is bounded using per-dimension minima and maxima computed from the GAE’s training embeddings, ensuring the search remains within a region likely to yield valid graphs.
The optimizer is configured with a population size of 64, leveraging parallel evaluations to enhance computational efficiency. To further optimize performance, a restart mechanism is implemented, allowing up to 10 restarts, with each limited to a maximum of 2\,000 function evaluations. The algorithm is based on the standard implementation provided by the \texttt{pycma} library~\cite{hansen2019pycma}.

\paragraph{Machine Learning classifier setup} To build classifiers, we label instances where RQAOA outperforms GW as 0, and those where GW outperforms RQAOA as 1. The dataset is split 70:30 into training and test sets. For the imbalanced 20-node dataset, we optimize for balanced accuracy~\cite{DBLP:conf/icpr/BrodersenOSB10}, whereas standard accuracy~\cite{scikit-learn} is used for the balanced 100-node case. We also conduct 10-fold cross-validation over 100 generations, evaluating 60 candidate models per generation. To further investigate RQAOA's reliability as a high-performance heuristic, we perform a secondary analysis using a modified labeling strategy. Here, instances are labeled as 1 if the ratio of GW to RQAOA performance is 0.96 or lower, indicating that RQAOA achieves at least a 4\% improvement over GW. All other instances are labeled as 0.

\paragraph{Reproducibility}
The full implementation is available in our Zenodo repository~\cite{supplementary}, along with supplementary materials for reproducibility and reference.

\subsection{Results}

\paragraph{Evaluation of the GAE}
During the training phase, we randomly save 60 graphs from a different set of 6\,000 graphs in each epoch, accumulating a total of 6\,000 graphs for in-depth evaluation. Our initial examination focuses on the 64-dimensional latent space structure, utilizing embeddings generated from these 6\,000 graphs. Initially, we compute several key features, as detailed in section~\ref{AppendixA} (see the supplementary material~\cite{supplementary}), prioritizing those that significantly influence the complexity and difficulty of the maximum cut problem. Research indicates that certain graph characteristics, especially eigenvalues, provide substantial insight into a graph's connectivity and complexity concerning the maximum cut problem~\cite{dunning2018works, moussa2020quantum}.



Fig.~\ref{fig:20_100node_log__secondlargesteigval} presents the log-transformed second-largest eigenvalues of the normalized Laplacian for both 20-node and 100-node graph instances. Each node in the plot corresponds to a cluster of graph embeddings identified through DBSCAN~\cite{ester1996density, schubert2017dbscan}, with edges indicating transitions or similarities between clusters in the latent space. We focus on the second-largest eigenvalue, rather than others, such as the spectral gap, because it provides more nuanced insight into structural properties such as loosely connected regions or modular communities, particularly when the largest eigenvalue is close to its upper bound, making the gap less informative. The color gradient across nodes highlights continuous changes in this spectral feature, reflecting varying levels of connectivity and complexity across clusters. This smooth variation demonstrates the smooth consistency of embeddings, validating the well-trained GAE’s suitability for generating new graphs for further analysis.

We continue to evaluate our trained GAE through a link prediction task, leveraging learned embeddings to capture graph topology. These embeddings facilitate adjacency matrix reconstruction, enabling predictions of potential node connections. Performance is measured using the macro area under the receiver operating characteristics curve (ROC-AUC) (0 to 1 scale, where 0.5 indicates random guessing and 1 denotes perfect prediction). Our initial assessment focuses on a GAE trained on 10-node Erdős–Rényi graphs (50\% probability) using a single A100 GPU with a batch size of 32. This setup yields a macro ROC-AUC of $0.933 \pm 0.004$ across 6\,000 graphs training graphs. Building on this promising result, we extend our analysis to larger graphs containing all graph models, obtaining macro ROC-AUC scores of $0.709 \pm 0.001$ for 20-node graphs and $0.611 \pm 0.001$ for 100-node graphs, using the same number of training samples.

\paragraph{Feature analysis through parallel coordinate plot}
\begin{figure}
    \centering
    \begin{subfigure}{\linewidth}
        \centering
        \includegraphics[width=\linewidth]{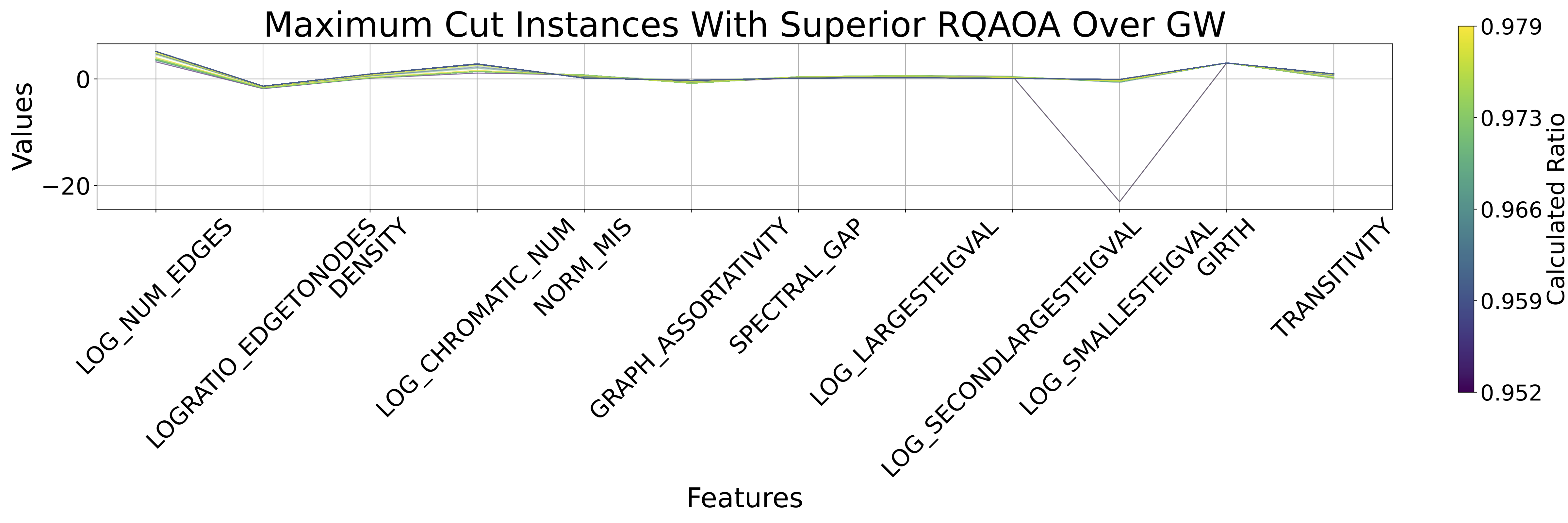}
        \includegraphics[width=\linewidth]{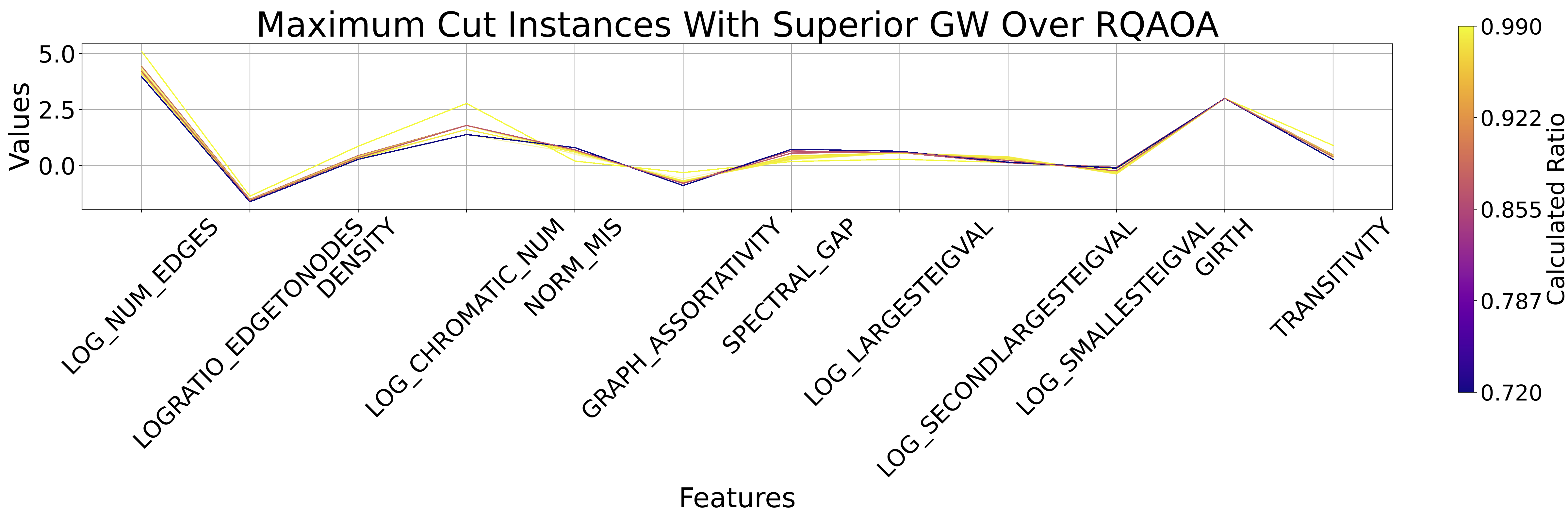}
        \caption{20-node hard maximum cut instances.}
        \label{fig: 20_node_features}
    \end{subfigure}
    \vspace{10pt}
    \begin{subfigure}{\linewidth}
        \centering
        \includegraphics[width=\linewidth]{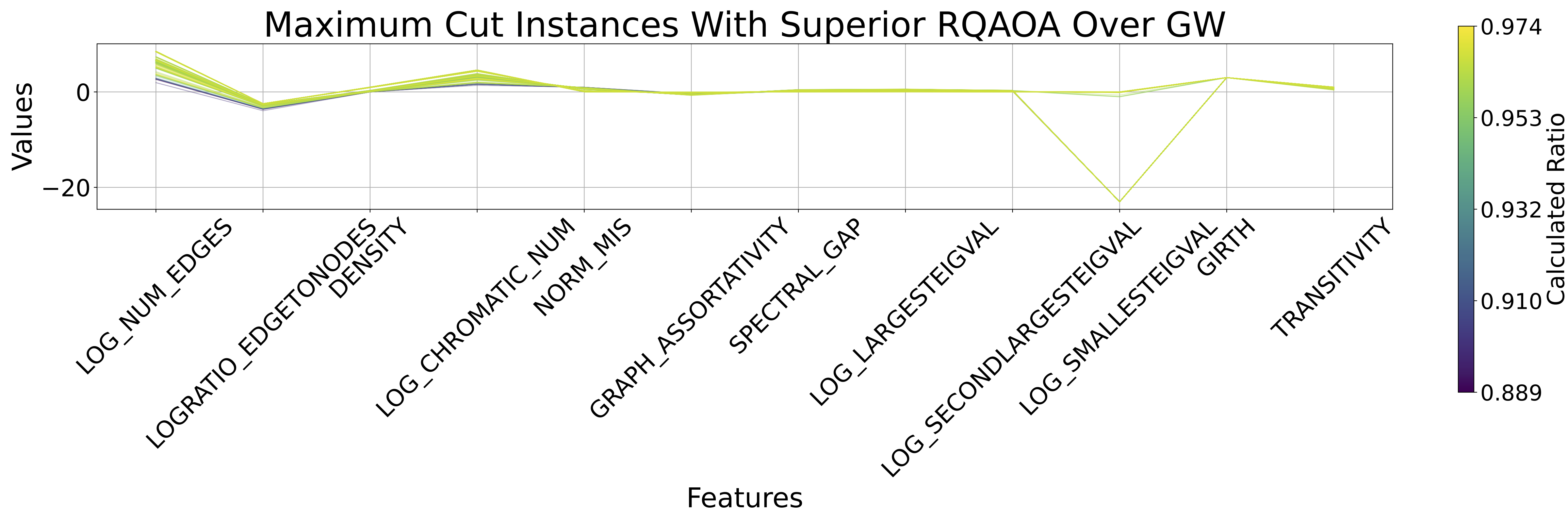}
        \includegraphics[width=\linewidth]{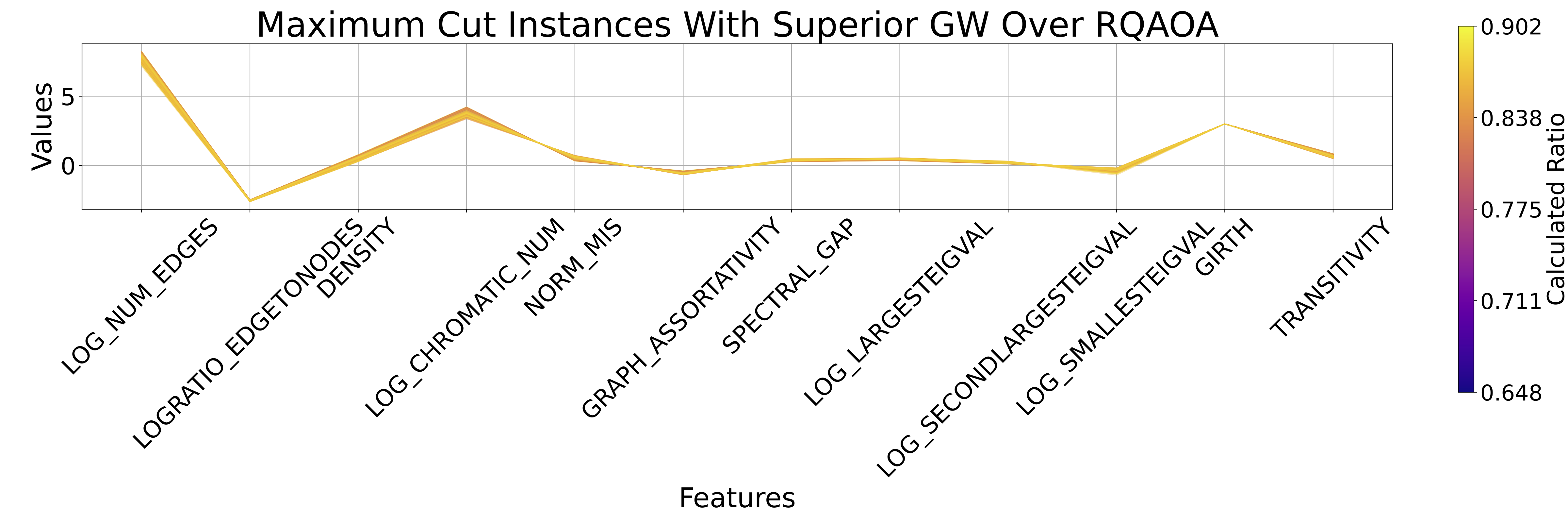}
        \caption{100-node hard maximum cut instances.}
        \label{fig: 100_node_features}
    \end{subfigure}

    \caption{Computed features of generated 20-node (top) and 100-node (bottom) hard maximum cut instances. Each pair of charts highlights cases where either the RQAOA or GW algorithm outperforms the other, with the color bar representing ratio values. Lower values indicate instances where one algorithm significantly outperforms the other.}
    \Description{Computed features of generated 20-node (top) and 100-node (bottom) hard maximum cut instances.}
    
\end{figure}

To ensure diversity, each CMA-ES run begins from different latent points within the latent space. For 20-node maximum cut instances, we generate 1\,519 instances where RQAOA outperformed GW and 1\,231 instances where GW exceeded the performance of RQAOA. Similarly, for 100-node instances, both algorithms outperform each other in 850 cases. To further analyze these instances, we examine various features (see section~\ref{AppendixA} in the supplementary material~\cite{supplementary}). Comparative results are visualized in Fig.~\ref{fig: 20_node_features} for 20-node graphs and Fig.~\ref{fig: 100_node_features} for 100-node graphs.

Fig.~\ref{fig: 20_node_features} presents a parallel coordinate plot comparing graph features for instances where either RQAOA or GW outperforms the other. Each line in the plot corresponds to a single graph instance, with the vertical position indicating the value of each feature for that instance. The top subplot highlights cases where RQAOA outperforms GW, with darker shades indicating greater superiority. The bottom subplot shows instances favoring GW, revealing that lower performance ratios (indicating stronger performance) appear more frequently when GW surpasses RQAOA. RQAOA-dominant instances exhibit a narrow performance ratio range (0.952–0.979), whereas GW-favored cases span a wider range (0.720–0.990). A key distinction emerges in \texttt{LOG\_SMALLESTEIGVAL}, where RQAOA-preferred graphs fall below -20, suggesting structural isolation. A representative case, featuring a mostly connected graph with one isolated node, illustrates this trend. Running 100 trials per algorithm on this instance, RQAOA outperforms GW in only 4 trials, with identical results in the rest, implying that GW’s inherent randomness may influence its comparative performance. Fig.~\ref{fig: 100_node_features} reveals a similar trend for 100-node graphs, indicating no significant differences in computed features compared to the 20-node case.


\paragraph{Feature analysis with machine learning pipeline}

To analyze the GW algorithm's role in generating hard maximum cut instances, we expand our feature set based on prior research~\cite{moussa2020quantum}. Incorporating six additional GW-related features (see section~\ref{AppendixB} in the supplementary material~\cite{supplementary}), we increase our total to 18 features within our machine learning pipeline. NP-hard features like the domination number are considered but excluded due to their computational complexity and minimal impact on model accuracy.

\begin{figure}
    \centering
    \begin{subfigure}{\linewidth}
        \centering
        \includegraphics[width=\linewidth]{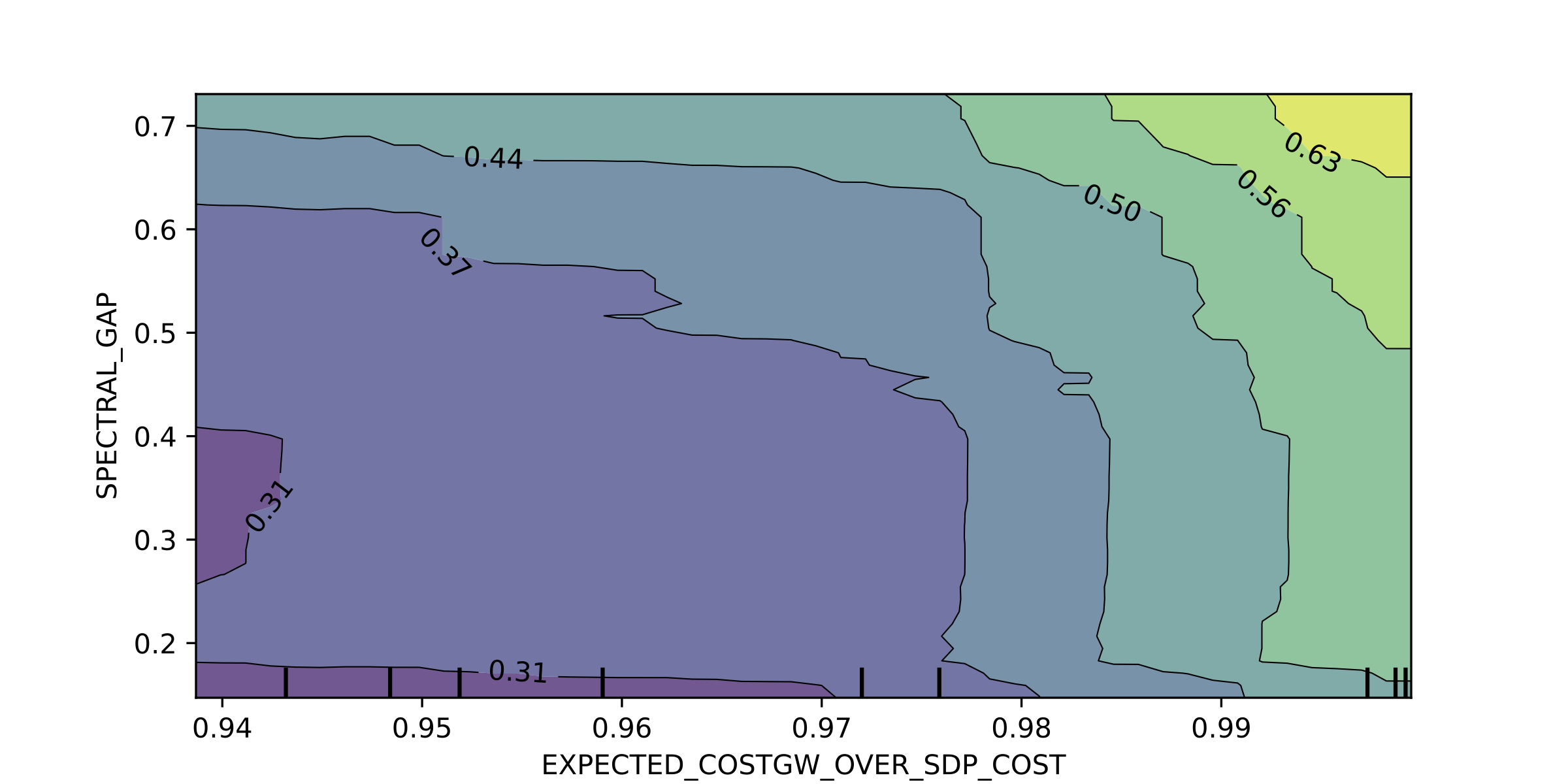}
        \caption{Partial dependence plot showing the impact of the top two features on 20-node maximum cut instances.}
        \label{fig:20node_PDP}
    \end{subfigure}
    \vspace{10pt}
    \begin{subfigure}{\linewidth}
        \centering
        \includegraphics[width=\linewidth]{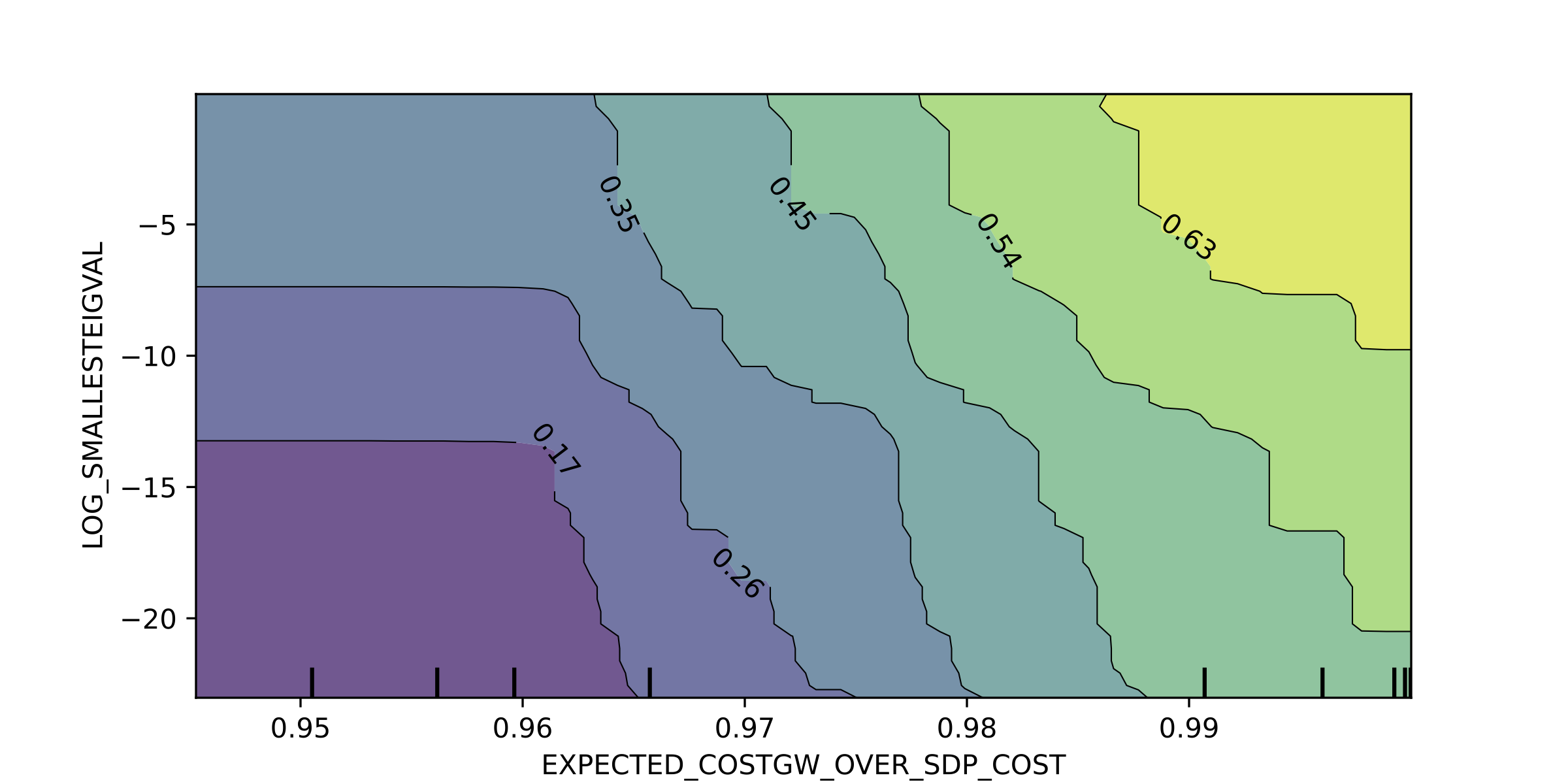}
        \caption{Partial dependence plot showing the impact of the top two features on 100-node maximum cut instances.}
        \label{fig:100node_PDP}
    \end{subfigure}
    \caption{Partial dependence plots showing the classifier’s response to the top two features of 20-node (top) and 100-node (bottom) maximum cut instances. The contour lines represent different probability levels, providing insight into feature influence.}
    \Description{Partial dependence plots illustrating the classifier’s behavior based on the top two features of 20-node (top) and 100-node (bottom) maximum cut instances.}
\end{figure}

In our 20-node experiments, we achieve an outstanding average balanced accuracy of $0.9988 \pm 0.0012$ across ten runs, each with a unique random data split. This result is obtained using an ExtraTreesClassifier~\cite{DBLP:journals/ml/GeurtsEW06} (parameters detailed in section~\ref{AppendixC} in the supplementary material~\cite{supplementary}). Key predictive features include \texttt{EXPECTED\_COSTGW\_OVER\_SDP\_COST} and \texttt{SPECTRAL\_GAP}, where the former evaluates GW's effectiveness, and the latter reflects graph spectral properties as shown in Table~\ref{tab:20nodefeatures} (see the supplementary material~\cite{supplementary}). The relevance of the GW algorithm's output in our analysis aligns with our focus on instances that are particularly challenging for maximum cut problems, influenced by the GW performance. The importance of eigenvalues also plays a crucial role, as seen in \texttt{LOG\_SECONDLARGESTEIGVAL} and \texttt{LOG\_LARGESTEIGVAL}, while \texttt{GIRTH} (shortest cycle length) contributes little to classifier performance. Further, the partial dependence plot visualizes the impact of two top features as shown in Fig.~\ref{fig:20node_PDP}. This plot demonstrates how these features influence the probability of predicting class 1, which in this study corresponds to instances where GW outperforms RQAOA. The plot reveals a clear trend: as the feature value changes, the probability of an instance being classified as class 1 increases, indicating a positive correlation between these features and the probability of challenging GW.

In the analysis of 100-node instances, we achieve perfect accuracy, recorded as $1.0000 \pm 0.0000$, using an ExtraTreesClassifier. The parameters for this classifier are outlined in section~\ref{AppendixC} (see the supplementary material~\cite{supplementary}). As indicated in Table~\ref{tab:100nodefeatures} (see the supplementary material~\cite{supplementary}), the feature \texttt{EXPECTED\_COSTGW\_OVER\_SDP} \newline \texttt{\_COST} stands out as the most significant, showing a much higher influence compared to other features. This finding is in line with our earlier results from the 20-node experiments, where four out of the five most influential features are similarly linked to the performance metrics of the GW algorithm. Consistent with previous observations, the feature GIRTH, which do not impact the classifier's decisions. The influence of the primary features is visualized in Fig.~\ref{fig:100node_PDP} through a partial dependence plot. It illustrates that as the values of \texttt{EXPECTED\_COSTGW\_OVER\_SDP\_COST} and \texttt{LOG\_SMALLESTEIGVAL} increase, there is a corresponding rise in the probability that an instance will perform better under GW compared to RQAOA.

To assess the significance of features, we further conduct a permutation important analysis. This involved randomly shuffling the values within columns of our dataset, focusing particularly on the top two features identified in both the 20-node and 100-node experiments. For each feature, we generate 10 different permutations of our original dataset. We then evaluate the impact of these permutations on the model's performance, which is initially trained on the unaltered dataset, using the same 10 seeds as the original setup. For the 20-node experiments, the permutation of the feature \texttt{EXPECTED\_COSTGW\_OVER\_SDP\_COST} results in a slight decrease in performance, with metrics changing by a $-0.0006\pm0.0018$. Interstingly, shuffling the \texttt{SPECTRAL\_GAP} feature leads to a minor performance increase, indicated by $0.0001\pm0.0017$. When both features are permuted simultaneously, the combined effect causes a more noticeable performance decrease of $0.0009\pm0.00177$. In the 100-node experiments, permutating the \texttt{EXPECTED\_COSTGW\_OVER\_SDP\_COST} feature lead to a more pronounced decline in performance, measured at $-0.0067\pm0.0048$. Permutation of the feature \texttt{LOG\_SMALLEST} \newline \texttt{\_EIGVAL} results in a smaller drop in performance, recorded as $-0.0004\pm0.0008$ on performance. Permuting both features together causes a decrease in performance of $-0.0012\pm0.0013$. Overall, we can conclude that the performance shifts upon permuting a single feature or the top two features slightly affect the model's performance.

To determine whether classifier performance could be sustained without the top two most informative features identified in our 20-node and 100-node experiments, we re-initiate the entire automated machine learning process excluding these features. In the 20-node experiments, this exclusion results in a minor decrease in performance, showing a decline of $0.0015\pm0.0022$ from the original baseline. Interestingly, in the 100-node experiments, performance remained unchanged, suggesting that the other features in the dataset still carry significant predictive power and are capable of compensating for the absence of the top two features. As indicated in the study~\cite{moussa2020quantum}, two features associated with GW performance are identified as the most significant factors for building the classifiers. Motivated to further explore this, we test whether classifiers built solely on GW-related features could maintain similar levels of performance. The results show a substantial drop in the 20-node experiment with accuracy falling to $0.9250 \pm 0.1541$, indicating that GW-related features alone are not sufficient at this scale. However, the 100-node experiment continues to achieve perfect accuracy of $1.0000 \pm 0.0000$. These findings suggest that while the GW-related features are insufficient on their own for the 20-node experiments, they provide substantial support in building the classifier with our 100-node instances.

\paragraph{RQAOA as a high-performing heuristic with machine learning pipeline}

\begin{figure}
    \centering
    \begin{subfigure}{\linewidth}
        \centering
        \includegraphics[width=\linewidth]{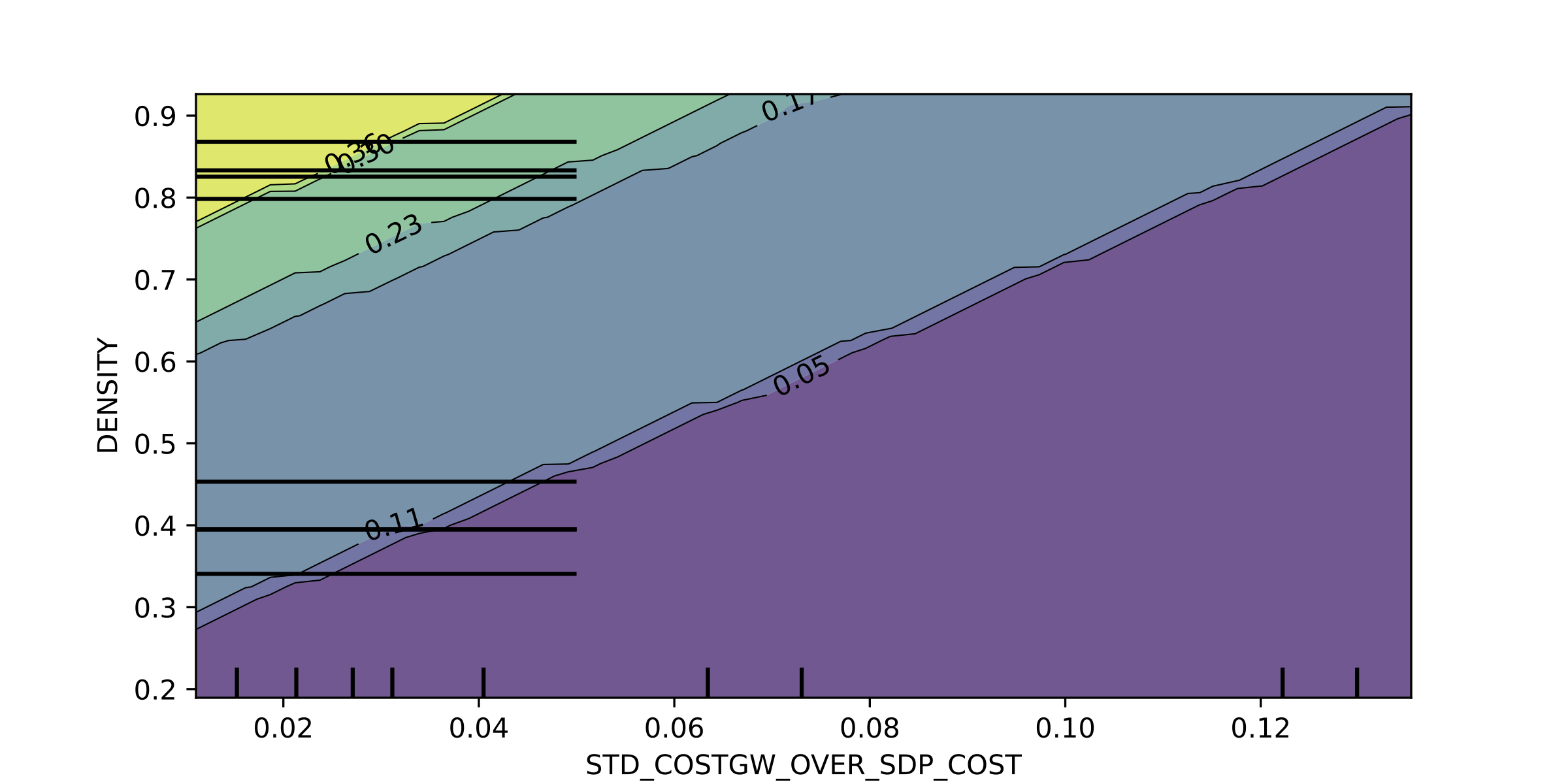}
        \caption{Partial dependence plot highlighting key features that distinguish significant RQAOA performance gains on 20-node maximum cut instances.}
        \label{fig:20node_PDP_high_performance}
    \end{subfigure}
    \vspace{10pt}
    \begin{subfigure}{\linewidth}
        \centering
        \includegraphics[width=\linewidth]{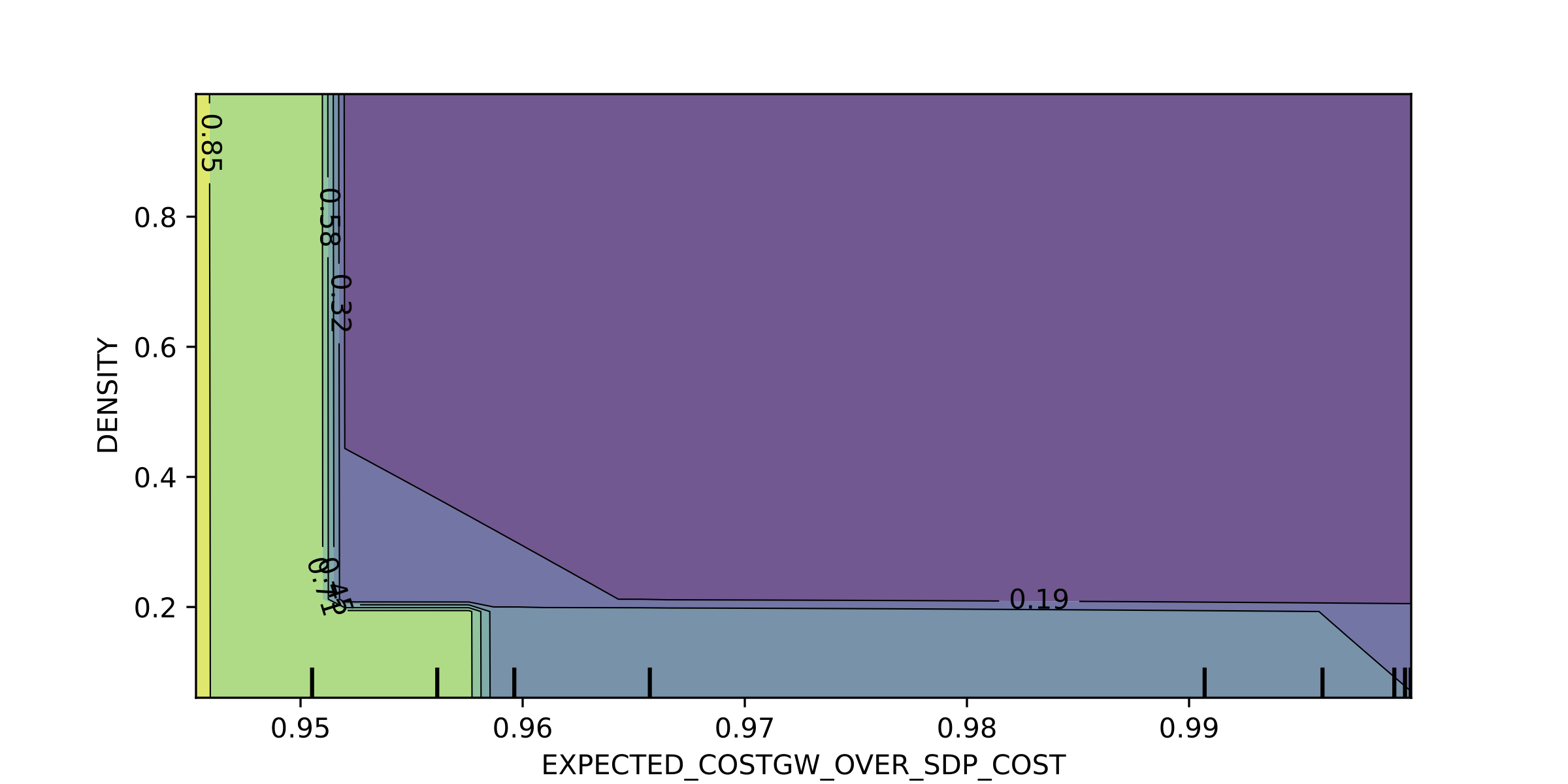}
        \caption{Partial dependence plot highlighting key features that distinguish significant RQAOA performance gains on 100-node maximum cut instances.}
        \label{fig:100node_PDP_high_performance}
    \end{subfigure}
    \caption{Partial dependence plots illustrating the classifier's behavior in identifying significant performance gains with RQAOA for 20-node (top) and 100-node (bottom) maximum cut instances. The plots highlight the influence of the top two features in distinguishing performance improvements, as determined by permutation importance.}
    \label{fig:20100node_PDP_high_performance}
    \Description{Partial dependence plots illustrating the classifier's behavior in identifying significant performance gains with RQAOA for 20-node (top) and 100-node (bottom) maximum cut instances. The plots highlight the influence of the top two features in distinguishing performance improvements, as determined by permutation importance.}
\end{figure}

After successfully applying our evolutionary approach to evolve hard maximum cut instances, a new question emerges: whether RQAOA consistently performs well? For our purposes, we seek to establish if RQAOA can achieve a performance level above 0.96, with a minimum advantage of 0.04 over the GW algorithm. 
We develop classifiers on both 20-node and 100-node instances that incorporate all the features previously discussed with achieved a balanced accuracy of $0.9826 \pm 0.0042$ on 20-node experiments and a $0.8343 \pm 0.0303$ in the 100-node experiment, maintaining the same experimental and analysis procedures as before. The models are detailed further in section~\ref{AppendixC} (see the supplementary material~\cite{supplementary}). Fig.~\ref{fig:20100node_PDP_high_performance} shows the partial dependent plot for the top two features in both 20-node and 100-node cases, respectively. The calculation of feature importance scores is based on permutation importance~\cite{scikit-learn}. The analysis of these plots reveals a clear pattern: low values in GW-related features are strongly associated with increased predicted probabilities of RQAOA achieving outstanding performance. Moreover, the significance of the \texttt{DENSITY} feature in influencing RQAOA's performance is evident. This analysis not only shows the impact of GW-related cost features but also emphasizes the critical role of \texttt{DENSITY} in determining the effectiveness of the RQAOA algorithm.

Table~\ref{tab:20nodefeatures_high_performing} and Table~\ref{tab:100nodefeatures_high_performing} (see the supplementary material~\cite{supplementary}) showcase the key features identified by the classifier for 20-node and 100-node maximum cut instances, respectively. Highlighted in these tables, the features emphasize the robust performance of RQAOA, validating its effectiveness as a high-performing heuristic through significant permutation importance.

We evaluate the performance of RQAOA against different thresholds to better understand our generated instances. Thresholds of 0.95 and 0.98 were initially considered, but discarded because, in each case, either the 20-node or 100-node instances failed to meet the criteria to be labeled as 1. At a threshold of 0.97, the average balanced accuracy achieved is $0.9795 \pm 0.0045$ for 20-node instances, which is comparable to the performance at a threshold of 0.96. \texttt{LOG\_CHROMATIC\_NUM} and \texttt{GRAPH\_ASSORTATIVITY} are identified as the most significant features according to permutation importance calculations. For 100-node instances, the performance improved notably at the 0.97 threshold, achieving an average balanced accuracy of $0.9214\pm0.0049$. This represents a substantial enhancement over the results obtained with the 0.96 threshold. The most influential features for this larger node size are identified as \texttt{PERCENT\_CLOSE3\_LOWER\_TRIANGULAR} and \texttt{GRAPH\_ASSORTATIVITY}. These results demonstrate that the model's effectiveness varies significantly with different threshold settings, with each threshold providing unique insights, especially in scenarios involving larger graph instances. This responsiveness underscores the need to thoughtfully choose performance benchmarks that align with the unique attributes and requirements of the dataset.

\section{Conclusion}

Given the current literature does not provide detailed guidelines for constructing hard maximum cut instances, our research addresses this gap. We focus on developing instances that highlight the distinctions between the quantum method RQAOA and the classical GW algorithm by effectively identifying maximum cut instances that pose significant challenges or are notably manageable for the RQAOA compared to the performance of the GW algorithm. Specifically, our work extends beyond situations where RQAOA merely surpasses GW, exploring instances where RQAOA achieves exceptionally high-quality outcomes and significantly outperforms GW. We employ an evolutionary approach using the classical CMA-ES algorithm, enhanced with a fitness function specifically tailored within the latent space of a well-trained Graph Autoencoder. 

Drawing on precedents that applied machine learning to analyze the performance of quantum methods~\cite{melnikov2019predicting, moussa2020quantum}, we adapt a comprehensive machine-learning pipeline. This approach examines a wide array of features, from those common to regular graphs and the spectrum of the Laplacian matrix to those associated with the relaxed solution of the semi-deﬁnite program in GW. Our feature analysis of the generated instances has further substantiated that RQAOA can offer substantial benefits over the GW algorithm. 

In predicting these advantages, the machine learning models we developed demonstrate strong results across both 20-node and 100-node instances. It confirms the necessity of both categories of features in analyzing instances where RQAOA significantly outperforms GW, reinforcing the comprehensive nature of the analytical framework. Our findings highlight \texttt{EXPECTED\_COSTGW\_OVER\_SDP\_} \newline \texttt{COST} as the most influential feature when RQAOA is not considered as a high-performing heuristic. This trend remains consistent in 100-node maximum cut instances when we consider RQAOA as a high-performing heuristic but does not hold for 20-node instances. Conversely, \texttt{DENSITY} plays a key role when RQAOA functions as a high-performing heuristic. Beyond these primary features, various other features also contribute to the machine learning classifiers, but \texttt{GIRTH} is found to have no impact in either experimental setting. 


Our study offers a fresh perspective on graph generation within network science, encouraging further exploration and advancements in this area. Future work can be explored from various perspectives. Our current experiments focus on graphs with 20 and 100 nodes, but we are actively working to scale up to larger instances, such as 400-node graphs. A key challenge in this expansion is the scalability of the PIGVAE model, originally designed for smaller graphs. Applying it to 400-node graphs causes severe memory bottlenecks. While we've reduced the model size to mitigate this, it may impact training performance.
Moving forward, we plan to test lightweight GNN architectures and use multi-GPU or distributed training to support larger models. For example, the use of sparse attention mechanisms is worth exploring, such as the implemented in Exphormers~\cite{DBLP:conf/icml/ShirzadVVSS23}. 
In addition to GNNs, diffusion models present a promising alternative for scalability, which can offer a more efficient scaling path.
As a next step, we will focus on 200–400 node graphs to evaluate and refine our approach before scaling further.
In parallel, we aim to enrich our analysis by incorporating additional classical heuristics beyond GW, offering a broader perspective on instance difficulty.

\section*{Acknowledgments}
This work has been supported by the Australian Research Council through grant FT200100536. 
YJP is supported by the `Quantum Inspire--the Dutch Quantum Computer in the Cloud' project [NWA.1292.19.194] of the NWA research program `Research on Routes by Consortia (ORC)', funded by Netherlands Organization for Scientific Research (NWO). The authors would like to thank Professor Chunhua Shen from Zhejiang University for insightful discussions on graph neural networks that contributed to the development of this research.
\bibliographystyle{ACM-Reference-Format}
\bibliography{abbrev}

\clearpage

\appendix
\renewcommand{\thetable}{A\arabic{table}}
\setcounter{table}{0}

\section{Graph Features}
\label{AppendixA}
The features of the instances utilized in this study are detailed as follows:
\begin{itemize}
    \item LOG\_NUM\_EDGES: The logarithm of the number of edges in the graph.
    \item DENSITY: The proportion of actual edges to the maximum possible number of edges (i.e., of a complete graph).
    \item LOGRATIO\_EDGETONODES: The natural logarithm of the edge-to-node ratio.
    \item LOG\_CHROMATIC\_NUM: The logarithm of the minimum number of colours required to colour the graph such that no two adjacent nodes share the same colour.
    \item NORM\_MIS: The size of the largest independent node-set, adjusted relative to the total number of nodes, showcasing the proportion of non-adjacent nodes.
    \item GRAPH\_ASSORTATIVITY: Gauges node connection similarity based on node degree, with values near 1 suggesting similar degree connections.
    \item SPECTRAL\_GAP: The difference between the largest and second-largest eigenvalues of the graph's normalized Laplacian matrix.
    \item LOG\_LARGESTEIGVAL: The logarithm of the highest eigenvalue from the normalized Laplacian matrix.
    \item LOG\_SECONDLARGESTEIGVAL:  The logarithm of the normalized Laplacian matrix's second-largest eigenvalue.
    \item LOG\_SMALLESTEIGVAL: The logarithm of the smallest non-trivial eigenvalue from the normalized Laplacian matrix.
    \item GIRTH: The length of the shortest loop in the graph.
    \item TRANSITIVITY: The likelihood ratio of nodes to form tightly connected clusters.
\end{itemize}

\section{GW-Related Features}
\label{AppendixB}
The features related to the GW algorithm are introduced as follows:
\begin{itemize}
    \item PERCENT\_CUT: The ratio of the optimal cut value derived from the semidefinite programming (SDP) solution to the overall number of edges in the graph.
    \item PERCENT\_POSITIVE\_LOWER\_TRIANGULAR: Quantifies the share of positive entries in the lower triangular section of the factorized matrix.
    \item PERCENT\_CLOSE1\_LOWER\_TRIANGULAR: The proportion of elements in the lower triangular of the factorized matrix that have absolute values under 0.1.
    \item PERCENT\_CLOSE3\_LOWER\_TRIANGULAR: The percentage of elements in the lower triangular of the factorized matrix with absolute values smaller than 0.001.
    \item EXPECTED\_COSTGW\_OVER\_SDP\_COST: The proportion of the GW approach's average cost (over 1,000 random trials) relative to the maximum theoretical cost as determined by the SDP method.
    \item STD\_COSTGW\_OVER\_SDP\_COST: Quantifies the standard deviation of the outcomes from 1,000 random implementations of GW, normalized by the best possible cost determined through the SDP method.
\end{itemize}

\section{Machine Learning Classifiers}
\label{AppendixC}
The models with hyperparameters constructed in the section of feature analysis with the machine learning pipeline are highlighted in the following:
\begin{itemize}
    \item The classifier developed using \texttt{TPOT} for instances involving 20 nodes uses an ExtraTreesClassifier, which comprises a collection of 100 decision trees. This ensemble employs the entropy criterion to determine the quality of splits within the trees. Each tree considers approximately 35\% of the features for node splitting. It also ensures that each leaf node has at least one sample, while at least nine samples are necessary to initiate further node splitting. The random\_state parameter is set to 42, ensuring reproducibility.
    \item The classifier built with \texttt{TPOT} for 100-node instances is implemented as an ExtraTreesClassifier that features bootstrapping. The entropy criterion is used to evaluate split quality across the trees. It also requires 80\% of the features to be considered for each tree split and a minimum of 19 samples for each leaf node. Additionally, a node split requires at least 5 samples to proceed. The ensemble includes 100 trees, and the random\_state parameter is fixed at 42.
\end{itemize}

The models with hyperparameters constructed in the section of RQAOA as a high-performing heuristic with the machine learning pipeline are highlighted in the following:
\begin{itemize}
    \item The classifier configured for 20-node instances using the \texttt{TPOT} library employs a sequential machine-learning pipeline. It begins with a StackingEstimator that deploys an MLPClassifier, characterized by a learning rate of 1.0 and a low regularization factor. This is complemented by another StackingEstimator, which utilizes an ExtraTreesClassifier. This classifier uses the entropy criterion method, avoids bootstrapping, and considers 65\% of the features during each split, all while adhering to predefined parameters for sample splits and minimum leaf size. The data is then normalized using a StandardScaler to ensure the features are optimally scaled. Finally, the process culminates with a GaussianNB classifier.
    \item The classifier configured using \texttt{TPOT} for 100-node instances also employs a sequential machine-learning pipeline. Initially, a StackingEstimator incorporating a GaussianNB classifier makes preliminary predictions that feed into later stages. Following that, a SelectFwe feature selector uses the f\_classif with an alpha of 0.012 to selectively maintain features. Subsequently, another StackingEstimator featuring a Decision Tree Classifier is used, which is tailored with the Gini index for node splitting, a cap of 7 on tree depth, a minimum of 9 samples per leaf, and at least 12 samples required for further splitting. A RobustScaler is then employed to adjust feature scales. The process concludes with a GaussianNB classifier.
\end{itemize}

\section{Important Features with Machine Learning Pipeline}
\label{AppendixD}

\begin{table}[H]
    \centering
    \begin{tabular}{rl}
        Important score & Feature\\
        \hline
        $0.1471 \pm 0.0059$ & EXPECTED\_COSTGW\_OVER\_SDP\_COST \\
        $0.1370 \pm 0.0084$ & SPECTRAL\_GAP \\
        $0.0769 \pm 0.0057$ & PERCENT\_CLOSE1\_LOWER\_TRIANGULAR \\
        $0.0684 \pm 0.0031$ & PERCENT\_POSITIVE\_LOWER\_TRIANGULAR \\
        $0.0665 \pm 0.0073$ & LOG\_SECONDLARGESTEIGVAL\\
        $0.0528 \pm 0.0042$ & LOG\_LARGESTEIGVAL \\
        $0.0527 \pm 0.0042$ & DENSITY \\
        $0.0521 \pm 0.0056$ & LOG\_NUM\_EDGES \\
        $0.0508 \pm 0.0051$ & PERCENT\_CLOSE3\_LOWER\_TRIANGULAR\\
        $0.0434 \pm 0.0068$ & LOGRATIO\_EDGETONODES \\
        $0.0431 \pm 0.0040$ & NORM\_MIS \\
        $0.0413 \pm 0.0036$ & GRAPH\_ASSORTATIVITY \\
        $0.0408 \pm 0.0042$ & TRANSITIVITY \\
        $0.0378 \pm 0.0027$ & STD\_COSTGW\_OVER\_SDP\_COST \\
        $0.0369 \pm 0.0049$ & PERCENT\_CUT \\
        $0.0310 \pm 0.0052$ & LOG\_SMALLESTEIGVAL \\
        $0.0215 \pm 0.0041$ & LOG\_CHROMATIC\_NUM \\
        $0.0000 \pm 0.0000$ & GIRTH \\
    \end{tabular}
    \caption{Important features with the classifier built on 20-node maximum cut instances.}
    \label{tab:20nodefeatures}
\end{table}
\begin{table}[H]
    \centering
    \begin{tabular}{rl}
        Important score & Feature\\
        \hline
        $0.3749 \pm 0.0104$ & EXPECTED\_COSTGW\_OVER\_SDP\_COST \\
        $0.1588 \pm 0.0196$ & LOG\_SMALLESTEIGVAL \\
        $0.1281 \pm 0.0052$ & PERCENT\_CLOSE1\_LOWER\_TRIANGULAR \\
        $0.0648 \pm 0.0081$ & PERCENT\_CUT \\
        $0.0618 \pm 0.0049$ & PERCENT\_CLOSE3\_LOWER\_TRIANGULAR\\
        $0.0454 \pm 0.0071$ & DENSITY \\
        $0.0322 \pm 0.0054$ & GRAPH\_ASSORTATIVITY \\
        $0.0239 \pm 0.0049$ & LOG\_LARGESTEIGVAL \\
        $0.0199 \pm 0.0085$ & LOG\_NUM\_EDGES \\
        $0.0165 \pm 0.0037$ & NORM\_MIS \\
        $0.0153 \pm 0.0036$ & TRANSITIVITY \\
        $0.0143 \pm 0.0026$ & LOG\_SECONDLARGESTEIGVAL\\
        $0.0126 \pm 0.0061$ & LOGRATIO\_EDGETONODES \\
        $0.0116 \pm 0.0026$ & LOG\_CHROMATIC\_NUM \\
        $0.0096 \pm 0.0030$ & SPECTRAL\_GAP \\
        $0.0061 \pm 0.0032$ & PERCENT\_POSITIVE\_LOWER\_TRIANGULAR \\
        $0.0043 \pm 0.0025$ & STD\_COSTGW\_OVER\_SDP\_COST \\
        $0.0000 \pm 0.0000$ & GIRTH \\
    \end{tabular}
    \caption{Important features with the classifier built on 100-node maximum cut instances.}
    \label{tab:100nodefeatures}
\end{table}

\section{Important Features with RQAOA as a High-performing Heuristic}
\label{AppendixE}

\begin{table}[H]
    \centering
    \begin{tabular}{rl}
        Important score & Feature\\
        \hline
        $0.0665 \pm 0.0641$ & STD\_COSTGW\_OVER\_SDP\_COST \\
        $0.0361 \pm 0.0382$ & DENSITY \\
        $0.0355 \pm 0.0389$ & NORM\_MIS \\
        $0.0352 \pm 0.0383$ & GRAPH\_ASSORTATIVITY \\
        $0.0347 \pm 0.0372$ & LOG\_LARGESTEIGVAL \\
        $0.0345 \pm 0.0376$ & LOG\_CHROMATIC\_NUM \\
        $0.0331 \pm 0.0344$ & TRANSITIVITY \\
        $0.0319 \pm 0.0201$ & SPECTRAL\_GAP \\
        $0.0180 \pm 0.0191$ & LOG\_SECONDLARGESTEIGVAL\\
        $0.0082 \pm 0.0051$ & LOG\_SMALLESTEIGVAL \\
        $0.0031 \pm 0.0034$ & PERCENT\_POSITIVE\_LOWER\_TRIANGULAR \\
        $0.0031 \pm 0.0039$ & LOG\_NUM\_EDGES \\
        $0.0028 \pm 0.0049$ & PERCENT\_CUT \\
        $0.0012 \pm 0.0025$ & LOGRATIO\_EDGETONODES \\
        $0.0010 \pm 0.0048$ & PERCENT\_CLOSE1\_LOWER\_TRIANGULAR \\
        $0.0002 \pm 0.0051$ & PERCENT\_CLOSE3\_LOWER\_TRIANGULAR\\
        $0.0000 \pm 0.0000$ & GIRTH \\
        $-0.0002 \pm 0.0029$ & EXPECTED\_COSTGW\_OVER\_SDP\_COST \\
    \end{tabular}
    \caption{Important features with the classifier built on 20-node maximum cut instances demonstrate RQAOA's high performance, underscored by its significant permutation importance.}
    \label{tab:20nodefeatures_high_performing}
\end{table}
\begin{table}[H]
    \centering
    \begin{tabular}{rl}
        Important score & Feature\\
        \hline
        $0.0243 \pm 0.0379$ & EXPECTED\_COSTGW\_OVER\_SDP\_COST \\
        $0.0166 \pm 0.0339$ & DENSITY \\
        $0.0114 \pm 0.0057$ & LOGRATIO\_EDGETONODES \\
        $0.0066 \pm 0.0110$ & NORM\_MIS \\
        $0.0036 \pm 0.0106$ & LOG\_SECONDLARGESTEIGVAL\\
        $0.0028 \pm 0.0047$ & LOG\_NUM\_EDGES \\
        $0.0023 \pm 0.0048$ & LOG\_LARGESTEIGVAL \\
        $0.0011 \pm 0.0020$ & PERCENT\_POSITIVE\_LOWER\_TRIANGULAR \\
        $0.0010 \pm 0.0052$ & LOG\_CHROMATIC\_NUM \\
        $0.0003 \pm 0.0027$ & STD\_COSTGW\_OVER\_SDP\_COST \\
        $0.0003 \pm 0.0004$ & SPECTRAL\_GAP \\
        $0.0001 \pm 0.0002$ & TRANSITIVITY \\
        $0.0000 \pm 0.0000$ & GIRTH \\
        $-0.0001 \pm 0.0004$ & PERCENT\_CLOSE1\_LOWER\_TRIANGULAR \\
        $-0.0001 \pm 0.0003$ & PERCENT\_CLOSE3\_LOWER\_TRIANGULAR \\
        $-0.0002 \pm 0.0004$ & PERCENT\_CUT \\
        $-0.0002 \pm 0.0008$ & GRAPH\_ASSORTATIVITY \\
        $-0.0153 \pm 0.0118$ & LOG\_SMALLESTEIGVAL \\
    \end{tabular}
    \caption{Important features with the classifier built on 100-node maximum cut instances demonstrate RQAOA's high performance, underscored by its significant permutation importance.}
    \label{tab:100nodefeatures_high_performing}
\end{table}

\end{document}